\documentclass[preprint,superscriptaddress,showpacs,preprintnumbers,amsmath,amssymb]{revtex4-1}

\usepackage{amsmath}
\usepackage{amssymb}
\usepackage{graphicx}
\usepackage{morefloats}
\usepackage[usenames,dvipsnames]{xcolor}
\usepackage{blkarray}
\usepackage{verbatim}
\usepackage{hyperref}
\usepackage{color}

\begin{document}

\title{Is Poker a Skill Game? New Insights from Statistical Physics}

\author{Marco Alberto Javarone}
\affiliation{Department of Mathematics and Computer Science, University of Cagliari, Cagliari (Italy)}
\affiliation{DUMAS - Department of Humanities and Social Sciences, University of Sassari, Sassari (Italy)}
\email{marcojavarone@gmail.com}

\begin{abstract}

During last years poker has gained a lot of prestige in several countries and, beyond to be one of the most famous card games, it represents a modern challenge for scientists belonging to different communities, spanning from artificial intelligence to physics and from psychology to mathematics.
Unlike games like chess, the task of classifying the nature of poker (i.e., as 'skill game' or gambling) seems really hard and it also constitutes a current problem, whose solution has several implications.
In general, gambling offers equal winning probabilities both to rational players (i.e., those that use a strategy) and to irrational ones (i.e., those without a strategy). Therefore, in order to uncover the nature of poker, a viable way is comparing performances of rational versus irrational players during a series of challenges. 
Recently, a work on this topic revealed that rationality is a fundamental ingredient to succeed in poker tournaments.
In this study we analyze a simple model of poker challenges by a statistical physics approach, with the aim to uncover the nature of this game. 
As main result we found that, under particular conditions, few irrational players can turn poker into gambling. Therefore, although rationality is a key ingredient to succeed in poker, also the format of challenges has an important role in these dynamics, as it can strongly influence the underlying nature of the game.
The importance of our results lies on related implications, as for instance in identifying the limits poker can be considered as a `skill game' and, as a consequence, which kind of format must be chosen to devise algorithms able to face humans.
\end{abstract}

\maketitle

\begin{figure*}
\centering
\includegraphics[width=1.0\textwidth]{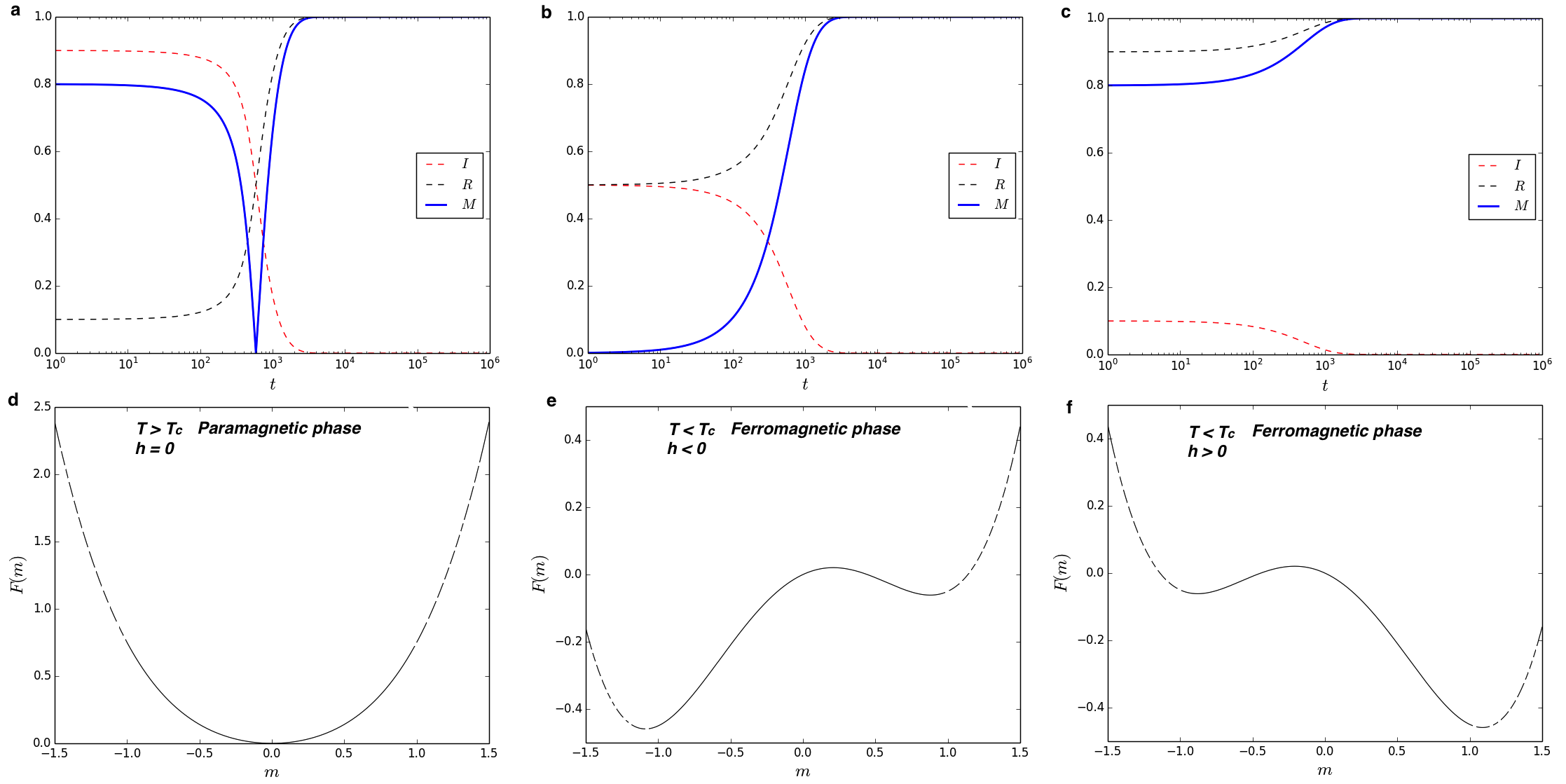}
\caption{\small From \textbf{a} to \textbf{c}: evolution of the system over time, according to the proposed $RIR$ model. The red dotted line represents the amount of irrational agents, the black dotted line represents the amount of rational agents, and the continuous blue line indicates the system magnetization $M$. \textbf{a} Results achieved by $\rho_r(0) = 0.1$. \textbf{b} Results achieved by $\rho_r(0) = 0.5$. \textbf{c} Results achieved by $\rho_r(0) = 0.9$. From \textbf{d} to \textbf{f}: Free energy $F$ as a function of the order parameter $m$, representing the behavior of the agent population on varying $\tau$ (i.e., the winning probabilities of rational and irrational agents). \textbf{d} Free energy, for $T > T_c$, related to the case $\tau = 0$, and with $\rho_r(0) = 0.5$. \textbf{e} Free energy, for $T < T_c$, related to the case $\tau < 0$. \textbf{f} Free energy, for $T < T_c$, related to the case $\tau > 0$. The order parameter $m$ has a domain defined in $[-1,+1]$, thus functions representing $F(m)$ are indicated by continuous lines inside the domain of $m$, and by dotted lines outside the domain. The dotted lines have been added in order to improve the quality of the pictorial representation of the minima of $F(m)$.\label{fig:mean_field}}
\end{figure*}

\begin{figure*}
\centering
\includegraphics[width=1.0\textwidth]{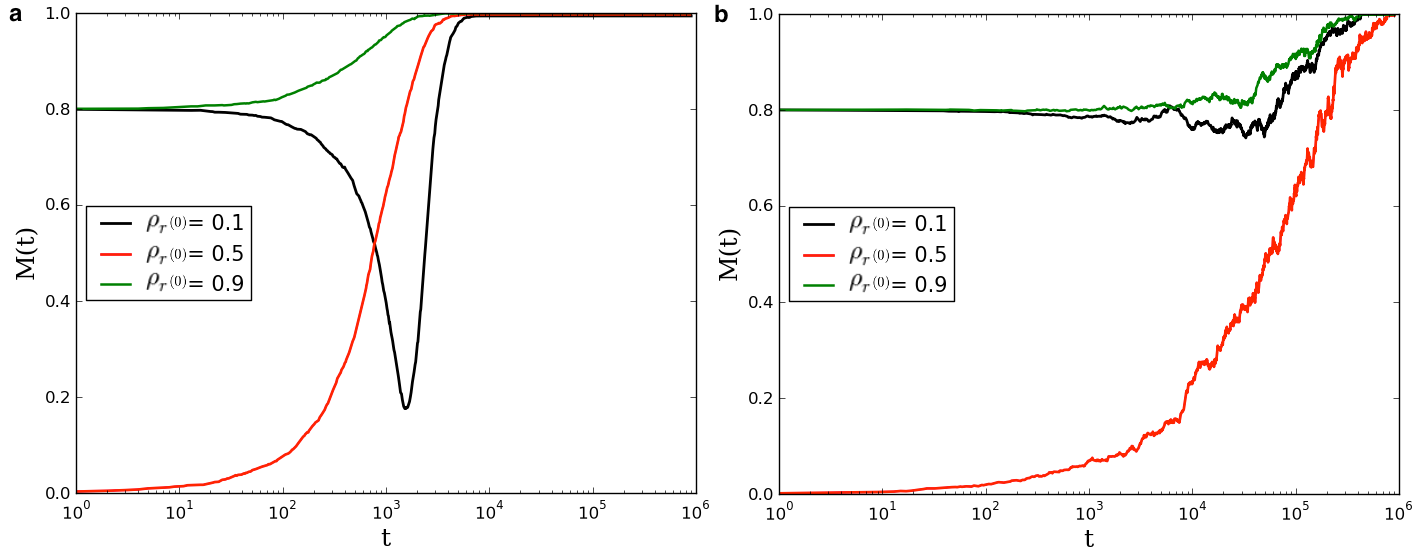}
\caption{\small Evolution of the magnetization on varying the initial density of rational agents $\rho_r(0)$. \textbf{a} Results achieved by implementing the case \textbf{a}: agents play full challenges. \textbf{b} Results achieved by implementing the case \textbf{b}: agents play single rounds. Results have been averaged over $100$ different simulation runs.\label{fig:magnetization}}
\end{figure*}

\begin{figure}
\centering
\includegraphics[width=0.48\textwidth]{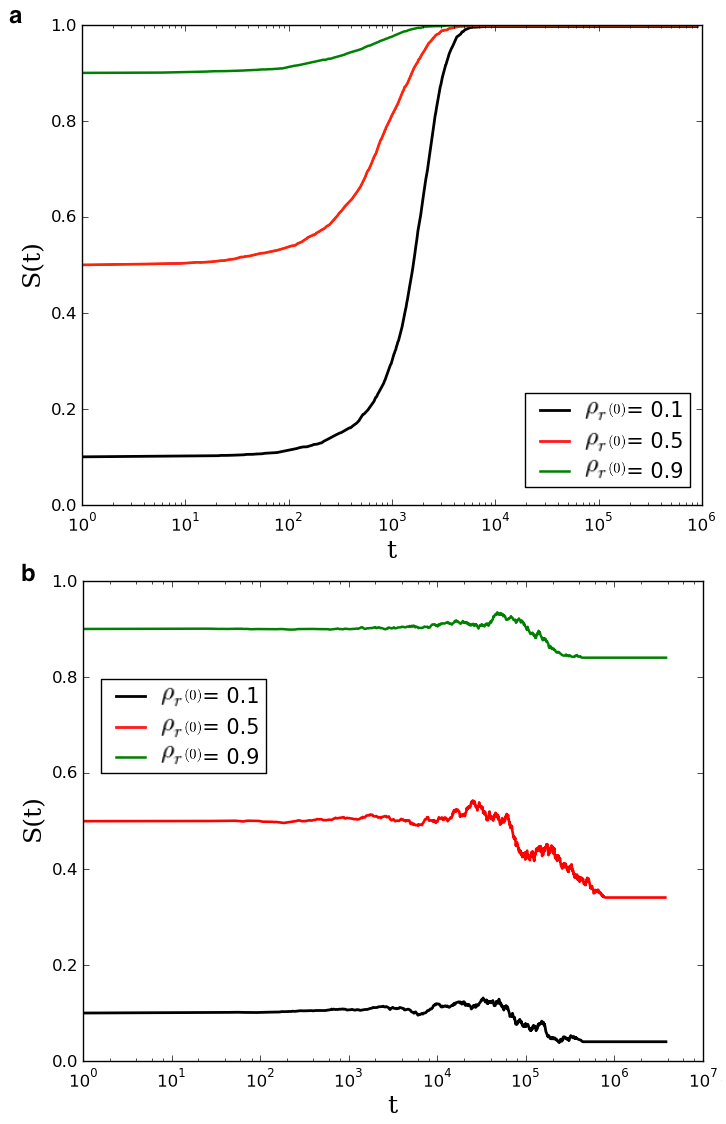}
\caption{\small Summation of states over time. \textbf{a} Results achieved by implementing the case \textbf{a}: agents play full challenges. \textbf{b}. Results achieved by implementing the case \textbf{b}: agents play single rounds. Results have been averaged over $100$ different simulation runs.\label{fig:summation}}
\end{figure}

Nowadays, social dynamics and modeling human behavior represent challenging topics for scientists belonging to different communities, e.g., artificial intelligence, physics, mathematics and social psychology.
Notably, the modern field of sociophysics~\cite{galam02} aims to investigate social and economic phenomena by a strongly interdisciplinary approach, mainly based on analytical and computational tools, coming from the framework of statistical physics~\cite{loreto01,barra01,barra02}.
Moreover, several social issues as opinion formation, information spreading and social behaviors, can be represented and studied by using agent-based models~\cite{rapisarda01,rapisarda02} often combined with the theory of networks~\cite{javarone01,tomassini01,javarone02,weron01}.
In this work, we analyze poker games (hereinafter simply poker) by the framework of statistical physics (see also ~\cite{sire01}). 
Poker represents one of the major challenges for artificial intelligence and mathematics ~\cite{bowling01,dahl01,teofilo01,seale01};for instance, it is worth to highlight that, in the recent study~\cite{bowling01}, the `heads up' limit poker (later described) has been solved from a game theory perspective. Furthermore, poker is a topic of interest also for psychologists, economists and sociologists~\cite{economist01} due to its wide diffusion over several countries.
One of the most controversial aspects of poker, caused by the utilization of money, is related to its nature, i.e., `skill game' or gambling. The related answer has not yet been solved~\cite{hannum01}, although it has a long list of implications~\cite{kelly01,cabot01}. 
Furthermore, all efforts made to define algorithms and strategies in the context of artificial intelligence are obviously based on the confident belief that computing skills are relevant to succeed in poker.
Therefore, our investigations aim to shed some light on the nature of this game.
In principle, there are several variants of poker, e.g., Texas Hold'em, Omaha, Draw, etc., each having its own rules. However, they all follow a similar logic: a number of cards is distributed among players, who in turn decide if to play or not, evaluating the possible combinations of their cards (called \textit{hand}) with those on the table. 
Since players cannot see the cards of their opponents, when they have to take an action (e.g., to bet money), poker is an imperfect information game, unlike others like chess where all players get all the system information simultaneously~\cite{dilemma01}.
It is worth to observe that the utilization of money makes the challenge meaningful, just because the underlying dynamics of poker are constituted by a series of bets. Hence, without money players would have no reasons to fold their \textit{hands}.
In general, there are two main formats for playing poker, i.e., tournament and `cash game'. The former entails players pay an entry fee that goes into the prize pool plus a fee to play, receiving an amount of chips. Then, top players share the prize pool. 
Instead, playing poker in the `cash game' format entails to use real money during the challenge. Therefore, in this last case, players can play until they have money and, although there are no entry fees to pay, a fraction of each pot is taxed (i.e., a small `rake' is applied).
In the work~\cite{javarone03}, the author defined a model for representing poker challenges, focusing on tournaments, in order to study the role of rationality. His main result was that the nature of poker does not depend on its rules but on the players's behavior, then identifying rationality as a key ingredient to succeed.
Hence, since the human behavior has such important role in poker, we perform further investigations on this direction, but considering the `cash game' format.
\newline
\newline
\indent Let us now briefly recall the model described in~\cite{javarone03} and summarize the main achievements. This model represents `heads-up' challenges, i.e., challenges that involve two players at a time. Players can be rational or irrational. The former move (e.g., bet and fold) by using the Sklansky table~\cite{sklansky01} as reference, whereas the latter play randomly.
It is worth to note that, for the sake of simplicity, each round is composed of only one betting phase (instead, in real scenarios, usually there are more phases~\cite{sklansky01}).
Numerical simulations showed that, under these conditions, rational players win a challenge against irrational players with probability $\pi^w_r \sim 0.8$.
Hence, a rational player is supposed to win about three consecutive challenges ($W=3$) against an irrational one.
As a consequence, since `heads-up' tournaments have a tree-like structure, the final winner is a rational player when the number of total participants $N$, regardless of their behavior, is $N \le 2^W$. 
After analyzing poker tournaments by different conditions (e.g., also allowing rationals to change behavior), the author~\cite{javarone03} states that the nature of poker depends on the players' behavior, but not on its rules.
\newline
\newline
\indent Here, we focus our attention on the `cash game' format. It is important to observe that each `heads-up' challenge can last from one to several rounds, in principle depending on the amount of money opponents have available.
Moreover, even after a single round one player can leave the table (i.e., ending the challenge) with her/his remaining money.
In order to study this scenario, we consider a population of agents that interact by the dynamics of the classical voter model~\cite{ligget01}. In so doing, each agent has a state that represents its behavior (i.e., rational or irrational) and, at each time step, two randomly chosen agents interact, i.e., they play a poker challenge. Notably, we map agent states as follows: $\sigma = +1$ for rational agents and $\sigma = -1$ for irrational agents.
Furthermore, we assume that a rational player wins a full challenge against an irrational one with the probability $\pi^w_r$ defined in~\cite{javarone03}. 
Therefore, the stochastic process of a poker challenge, involving players $x$ and $y$, is reduced to a coin flip with winning probabilities
\begin{equation}\label{eq:winning_prob}
\begin{cases} \pi^w_x = 0.8 & \mbox{if } (x != y  \mbox{ and } \sigma_x = +1)\\ 
\pi^w_x = 0.5 & \mbox{if } (x=y) \end{cases}
\end{equation}
 \noindent and $\pi^w_y = 1 - \pi^w_x$. Then, according to voter model-like dynamics, after each interaction the loser assumes the state (i.e., the behavior) of the winner.
Since, as said before, a cash game challenge can last from $1$ to $n$ rounds, we have two limit cases: 
\begin{enumerate}
\item \textbf{a}) $n = \infty$: an interaction corresponds to a full challenge, as after an infinite number of rounds one player prevails;
\item \textbf{b}) $n = 1$: an interaction corresponds to only one round.
\end{enumerate}
In both cases, agents start a new challenge always with the same amount of money (called `starting stack'), regardless of their previous results. Furthermore, in the cash game format the minimal amount of a bet (called `big blind') does not change over time. In the proposed model, we set the `starting stack' to $10000$ and a `big blind' to $100$ so, considering the dynamics of case \textbf{a}, the ratio between these two parameters only affects the length of challenges (see also ~\cite{javarone03}), while in the case \textbf{b} it has no influences. 
It is worth to highlight that both \textbf{a} and \textbf{b} refer to real scenarios. In particular, the case \textbf{b} represents the so called `rush poker', available in several online platforms. Moreover, we highlight that for the case \textbf{b} Eq~\ref{eq:winning_prob} cannot be used, since it holds only for a full challenge.
Anyway, also for the case \textbf{b}, it would be possible to derive the winning probabilities for rational agents by analyzing the Sklansky table~\cite{sklansky01}.
\newline
\newline
Now, we introduce a mean field approximation~\cite{barra03} of the proposed model considering, in particular, the case \textbf{a} as the winning probability of rational agents is defined (see equation~\ref{eq:winning_prob}).
Since agents can change state over time, i.e., from rational to irrational and vice versa, the following equations describe the dynamics of the population
\begin{equation}\label{eq:rir_model}
\begin{cases}\frac{d\rho_r(t)}{dt} = a \cdot \rho_r(t) \cdot \rho_i(t) - b \cdot \rho_i(t) \cdot \rho_r(t)\\ 
\frac{d\rho_i(t)}{dt} =  b  \cdot \rho_i(t) \cdot \rho_r(t) - a \cdot \rho_r(t) \cdot \rho_i(t)\\
\rho_r(t) + \rho_i(t) = 1
\end{cases}
\end{equation}

\noindent with $\rho_r$ and $\rho_i$ density of rational and irrational agents, respectively. Parameters $a$ and $b$ represent the winning probabilities of each species (i.e., rational and irrational), then $a = \pi^w_r$ and $b = 1 - \pi^w_r$. 
In so doing, we are studying the system by a compartmental approach, as in $SIS$-like models adopted in computational epidemics~\cite{anderson01}. As result we achieve a $RIR$ model (i.e., $Rational \to Irrational \to Rational$) whose solutions, computed by integrating between $0$ and $\frac{t}{N}$ (as we are considering an asynchronous dynamics) are
\begin{equation}\label{eq:rir_solution}
\begin{cases} \rho_r(t) = \rho_r(0) \cdot e^{\frac{\rho_i(t) \tau}{N} t}\\ 
\rho_i(t) = \rho_i(0) \cdot e^{-\frac{\rho_r(t) \tau}{N} t}
\end{cases}
\end{equation}
\noindent with $\tau = (a - b)$ which justifies the minus in the exponent of the second equation in the system~\ref{eq:rir_solution}.
Panels \textbf{a,b,c} of figure~\ref{fig:mean_field} show $\rho_r(t)$ and $\rho_i(t)$ over time, according to Equations~\ref{eq:rir_solution}, and the value of the system magnetization $M$ defined as follows~\cite{mobilia01}
\begin{equation}\label{eq:magnetization}
M = \frac{|\sum_{i=1}^{N}\sigma_i|}{N}.
\end{equation}
Now, we focus our attention on the behavior of the system for $t \to \infty$: for $\tau > 0$, we obtain $\rho_i \to 0$ and $\rho_r = 1$, whereas for $\tau < 0$, the opposite happens, i.e., $\rho_i \to 1$ and $\rho_r = 0$. In the case $\tau = 0$, as $t \to \infty$, the final equilibrium state corresponds to the co-existence of rational and irrational agents, whose final values are $\rho_r(0)$ and $\rho_i(0)$, respectively. 
It is worth to note that, this third equilibrium can be obtained also by solving the system of equations~\ref{eq:rir_model} by setting $\frac{d\rho_r(t)}{dt} = 0$ and $\frac{d\rho_i(t)}{dt} = 0$ as, in doing so, we find $a = b$. 
Remarkably, the system behavior can be described by the Curie-Weiss model~\cite{barra03} (hereinafter CW). Notably, the agent population for $\tau = 0$ behaves as a spin system at $T > T_c$ (i.e., whose temperature is greater than the Curie or `critical' temperature), where the system equilibrium corresponds to a disordered phase.
Instead, in the cases $\tau > 0$ and $\tau < 0$ the population behaves as a spin system at $T < T_c$, having two possible equilibria, both corresponding to an ordered phase.
Therefore, for $\tau$ equal to zero, the agent population is in a paramagnetic phase while, for values of $\tau$ greater or lower than zero, it is in a ferromagnetic phase.
A possible solution, to analytically describe the agent population by using the CW model, can be devised by mapping $\tau$ to an external magnetic field $h$.
In doing so, we can study the system equilibria by analyzing the free energy $F$. Notably, according to the Landau mean field theory~\cite{huang01}, the free energy $F$, as a function of the order parameter $m$, can be defined as
\begin{equation}\label{eq:landau}
F(m) = -hm + \alpha(T)m^2 + \frac{u}{2}m^4
\end{equation}
It is strongly important to note that, in a classical spin system, the magnetization $m$ spans from $-1$ to $+1$, and it is computed as follows
\begin{equation}\label{eq:classical_magnetization}
m = \frac{\sum_{i=1}^{N} \sigma_i}{N}
\end{equation}
\noindent with $N$ number of spins, whose value can be $\sigma = \pm 1$. Therefore, the definition of the magnetization $M$ in equation~\ref{eq:magnetization} (usually adopted in the context of opinion dynamics) and that of $m$, defined in equation~\ref{eq:classical_magnetization}, coincide for less than the absolute value adopted in the former, i.e., $M = |m|$.
Now, the extrema of Eq.~\label{eq:landau} can be found by computing solutions of $\frac{dF}{dm} = 0$.
In particular, for $h = 0$, the two solutions are $m = (0, \pm \sqrt{-\frac{\alpha(T)}{u}})$. 
Here, $\alpha = c  \frac{T - T_c}{T_c}$, with $c$ small positive constant, and $u$ positive parameter, that we set to $u = \frac{1}{2}$. 
It is worth to observe that $\alpha(T)$ becomes zero as $T \to T_c$, i.e., as the system temperature approaches the `critical' temperature.
For the paramagnetic phase (i.e., $T > T_c$) we can set $\alpha = \frac{1}{2}$, while for the ferromagnetic phase we can set $\alpha = -\frac{1}{2}$. Then, Equation~\ref{eq:landau} becomes
\begin{equation}\label{eq:landau_modify}
F(m) = -hm \pm \frac{m^2}{2} + \frac{m^4}{4}
\end{equation}
\noindent where the sign of the second term depends on the system temperature, i.e., positive for $T > T_c$ and negative for $T < T_c$.
We recall that, in the paramagnetic phase, there is a unique minimum of free energy for $m = 0$ ---see panel \textbf{d} of figure~\ref{fig:mean_field}.
On the other hand, in the ferromagnetic phase there are two minima of free energy, which correspond to $\pm 1$. Remarkably, for values of $h$ greater or lower than zero, one of the two minima becomes an absolute minimum of free energy: for $h > 0$ the minimum corresponding to $m = +1$ becomes deeper than that corresponding to $m = -1$, while the opposite happens for $h < 0$ (i.e., $F(-1) < F(+1)$)  ---see panel \textbf{e,f} of figure~\ref{fig:mean_field}.
Then, we can study by an analytical approach the outcomes of the proposed model in the case \textbf{a}, on varying the value of the winning probability of rational agents $\pi^w_r$.
We observe that the same approach cannot be used to analyze the case \textbf{b}, of the proposed model, as the winning probabilities are not defined a priori as for the case \textbf{a}.
Therefore, the expected behavior of the case \textbf{b} is more complex, and we hypothesize that its dynamics can show the presence of bifurcations. Notably, since the Sklansky table~\cite{sklansky01} suggests to play usually with a small set of \textit{hands}, many rounds will be won by irrational agents due to several `fold' actions performed by rational ones. 
As result, it is possible that even for a high initial density of rational agents, sometimes few irrational agents prevail then, for the same $\rho_r(0)$, the final state of the population can be both $+1$ (rational) and $-1$ (irrational), i.e., a bifurcation emerges.
\newline
\newline
\indent The proposed model is now studied by numerical simulations. In particular, we consider populations of different size, from $N = 100$ to $N = 1000$ agents, and we perform for each case $100$ simulation runs.
Since we aim to compare performances of rational versus irrational agents, each simulation lasts until all agents converge to the same behavior (i.e., state). As shown before for the $RIR$ model, it is possible to analyze the evolution of the system, for different initial densities of rational agents, by studying the magnetization $M$.
We recall that the value of $M$, according to Equation~\ref{eq:magnetization}, ranges between $0$ and $1$ (i.e., $0\leq M\leq1$). 
When $M \sim 0$, the system is in a disordered phase as there is the same amount of agents in the two states, whereas as $M \to 1$ the system reaches an ordered phase, characterized by the presence of a prevailing state ($\sigma = +1$ or $\sigma = -1$).
Figure~\ref{fig:magnetization} shows the magnetization over time, achieved in the two considered cases (i.e.,\textbf{a} and \textbf{b}).
Notably, both varying the density $\rho_r(0)$ and considering the two cases, the agent population always converges to the same state, in full accordance with the analytical predictions shown in Figure~\ref{fig:mean_field} (for the case \textbf{a}). 
Before to proceed with further analyses, it is worth to spend few words in order to explain why the magnetization $M$ in the panel \textbf{a} of figure~\ref{fig:magnetization} does not reach zero as in the related analytical solution (see panel \textbf{a} of figure~\ref{fig:mean_field}), i.e., for $\rho_r(0) = 0.1$. Notably, recalling that values of $M$ achieved in numerical simulations have been averaged over different runs, at each single attempt the time step $t$ corresponding to $M=0$ may vary, as we are dealing with a stochastic process, hence by averaging all results the average minimum value is not zero.
At this point, it is worth to investigate the final population state ($\Sigma$), in order to know whether, after all challenges, agents play rationally (i.e., $\Sigma = +1$) or not (i.e., $\Sigma = -1$).
Thus, we analyze the amount of rational agents over time $S(t)$, for different initial densities $\rho_r(0)$ ---see figure~\ref{fig:summation}.
Remarkably, since values of $S(t)$ are averaged over different simulation runs, and by knowing that at each attempt the population reaches an ordered phase, we may derive the probability $P^w_r$ that rational agents prevail on irrational ones on varying $\rho_r(0)$. 
Notably, these winning probabilities have been computed for different values of $\rho_r(0)$ from $0$ to $1$, focusing on small values close to $0$ (e.g., $0.0033$, $0.01$, $0.05$) for the case \textbf{a} and on high values close to $1$ (e.g., $0.97$, $0.98$, $0.99$) for the case \textbf{b}. 
The main reason to explore in particular low $\rho_r(0)$ for the first case and high $\rho_r(0)$ for the second case lies in the fact that, observing figure~\ref{fig:summation}, we found that rational agents easily prevail playing full challenges (i.e., \textbf{a}) against irrational agents that, in turn, prevail many times playing single rounds (i.e., \textbf{b}).
We want to highlight that results shown in panel \textbf{a} of figure~\ref{fig:summation} are completely in accordance with the analytical solution, as simulations of case \textbf{a} have been always performed with $\tau >0$ as $\pi^w_r = 0.8$ (see panel\textbf{f} of figure~\ref{fig:mean_field}); moreover, as we hypothesized before, a more complex behavior emerges in the case \textbf{b}.
\begin{figure}
\centering
\includegraphics[width=0.48\textwidth]{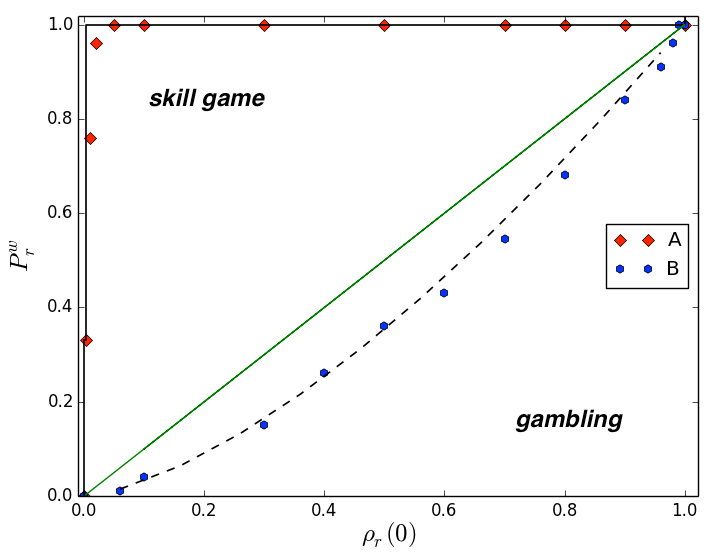}
\caption{\small Probability that rational agents prevail ($P^w_r$) on varying $\rho_r(0)$. In the legend, $A$ refers to the case \textbf{a} and $B$ to the case \textbf{b}. The two black lines (i.e., the dotted and the continuous one) refer to the computed fitting functions. The green continuous line separates the upper side of the plane, i.e., the `skill game' area, from the lower side, i.e., the gambling area.\label{fig:probability}}
\end{figure}
In particular, when agents play single rounds, the value of $M$ only increases up to $1$, but considering the summation $S(t)$, we observe that few irrational agents can sometimes turn into irrational the whole population even for a high initial density of rational agents.
Eventually, figure~\ref{fig:probability} further highlights the detected differences between the two considered scenarios. Notably, we computed fitness functions for both cases, identifying a simple step function for \textbf{a}, and the function 
\begin{equation}\label{eq:fitting_b}
P^w_r(\rho_r(0)) = \rho_r(0)^{3/2}
\end{equation}
\noindent for \textbf{b}. It is worth to recall that the function defined in equation~\ref{eq:fitting_b} allows to fit results of simulations (in the case \textbf{b}), and it has not been defined by the analytical approach.
On one hand, it is interesting to observe that in full challenges even the presence of only one rational agent can entails the transition to an ordered `rational' phase. On the other hand, when playing single rounds, rational agents prevail with a probability greater than $50\%$ only if $\rho_r(0) > 0.7$ hence, in our opinion, poker in this last case can be considered as gambling.
\newline
\newline
\noindent All these results confirm that classifying the nature of poker is a tricky task, as a lot of conditions must be considered in real scenarios. In particular, according to the proposed model, although in tournaments it seems rationality be a key ingredient to succeed~\cite{javarone04}, in the cash game format it may be sometimes appropriate to associate poker to gambling.
Moreover, considering all risks of poker in the cash game format (see~\cite{javarone04}), we think both players and scientists working on poker be aware of our results.
A further important point to discuss, before to conclude, is related to the validity of out model in real scenarios. Notably, although it would be extremely interesting to compare outcomes of the proposed model with real data, this is not possible as no similar datasets exist. 
Anyway it may be possible to evaluate if a player is adopting mainly a random strategy or a rational one as, according to the rules~\cite{sklansky01}, often players have to show their \textit{hands} after the round to discover who is the winner. 
Finally, we deem the importance of our results lies on related implications. Notably, we found that not only the player's behavior but also the format of poker must be considered when classifying the nature of this game, showing that there are well defined limits poker can be considered as a `skill game'.
\acknowledgments
MAJ is extremely grateful to Adriano Barra for all priceless suggestions. Furthermore, the author would like to thank Fondazione Banco di Sardegna for supporting his work.

\end{document}